\documentclass{article}

\usepackage[english]{babel}

\usepackage[letterpaper,top=2cm,bottom=2cm,left=3cm,right=3cm,marginparwidth=1.75cm]{geometry}

\usepackage{booktabs}
\usepackage{verbatim}
\usepackage{amsmath}
\usepackage{graphicx}
\usepackage{graphicx}
\usepackage{subcaption}
\usepackage{appendix}
\usepackage{placeins}
\usepackage[colorlinks=true, allcolors=blue]{hyperref}
\usepackage{geometry}
\usepackage{longtable}
\usepackage{amssymb}
\usepackage{pifont}
\usepackage{authblk}

\title{Anatomy of Elite and Mass Polarization in Social Networks}

\author[1]{Ali Salloum}
\author[2]{Ted Hsuan Yun Chen}
\author[3]{Mikko Kivelä}
\affil[1, 3]{Department of Computer Science, Aalto University}
\affil[2]{Department of Environmental Science and Policy, George Mason University}

\begin{document}

\maketitle

\begin{abstract}
\noindent

In the political arena of social platforms, opposing factions of varying sizes show asymmetrical patterns, and elites and masses within these groups have divergent motivations and influence, challenging simplistic views of polarization. Yet, existing methods for quantifying polarization reduce division to a single value, assuming uniform distribution of polarization online. While this approach can confirm the observed increase in political polarization in many societies, it overlooks complexities that could explain this phenomenon. Notably, opposing groups can have unequal impacts on polarization, and the literature shows division between elites and the masses is a critical factor to consider.

We propose a method to decompose existing polarization measures in order to quantify the role of groups, determined by these distinct hierarchies, in the total polarization value. We applied this method to polarized topics in the Finnish Twittersphere surrounding the 2019 and 2023 parliamentary elections. Our analysis reveals two key insights: 1) The impact of opposing groups on observed polarization is rarely balanced, and 2) while elites strongly contribute to structural polarization and consistently display greater alignment across various topics, the masses have also recently experienced a surge in issue alignment, a stronger form of polarization. 

Our findings suggest that the masses may not be as immune to an increasingly polarized environment as previously thought. This research provides a more nuanced understanding of polarization dynamics, offering potential insights into its underlying mechanisms and evolution.
\end{abstract}

\section{Introduction}

In an era of increasing political polarization \cite{silva2018populist, iyengar2019origins, bottcher2020great, KEKKONEN2021102367, falkenberg2023affective, karjus2024evolving, 10.1145/3589334.3645651}, the ability to dissect and understand the intricate structures within social systems has never been more crucial. The proliferation of social media platforms has transformed the landscape of political discourse, creating vast digital arenas where polarization can be observed, measured, and analyzed in unprecedented detail. 

Current approaches to measuring polarization often reduce complex social dynamics to a single dimension, failing to expose the hidden patterns behind the observed societal divisions. While these methods have been successfully applied to report polarization trends online \cite{GARIMELLA2018}, they fall short of capturing the full picture. This oversimplification obscures the multifaceted nature of polarization, where different groups and their members may contribute to and experience division in distinct ways. 

This work proposes a method that utilizes existing polarization scores and network representations of social systems to measure two key types of polarization—\textit{structural polarization} and \textit{issue alignment}—separately for the elites and masses in polarized systems. Our approach not only unveils significant imbalances between polarized groups but also highlights the disproportionate role of elites in shaping polarized environments.

The distinction between elite and mass polarization is crucial for understanding the true nature of political division in modern democracies. While headlines and public discourse often paint a picture of a deeply divided populace, this perception may be skewed by the outsized influence of political elites in media narratives. By conflating elite and mass polarization, we risk misdiagnosing the health of our democratic systems and implementing misguided solutions. Separating these measurements allows us to discern whether the apparent chasm in political attitudes truly reflects widespread societal division or if it predominantly exists among a small, albeit influential, segment of the population.

In addition, the relationship between political elites and the public is characterized by substantial asymmetry, with elites often serving as influential cue-givers \cite{levendusky2010clearer, diermeier2019partisan, green2020covid, skytte2021dimensions, van2021elite} who shape mass attitudes and behaviors \cite{berinsky2007assuming,
levendusky2010clearer,kousser2018influence, merkley2018party, merkley2021party, green2020covid, alley2023elite}. This \textit{elite signaling} phenomenon, where followers adjust their stances in response to elite ideological divergence or out-group hostility \cite{berinsky2007assuming, banda2018elite, merkley2018party, fine2023negativity}, has been linked to political polarization \cite{skytte2021dimensions, back2023elite}. Traditionally, elites have been viewed as more ideologically coherent and predictable than the general public, with some theories (e.g. \cite{layman2002party, kinder2017neither}) even suggesting that masses remain resistant to elite polarization while opposing elite groups grow increasingly distant and internally homogeneous \cite{robison2016elite}. However, recent research complicates this narrative, revealing potential ideological realignment among the masses \cite{levendusky2010clearer, kozlowski2021issue}. Our method enables tracking of these complex dynamics on online platforms, offering a real-time lens into whether and how elite-mass polarization patterns are evolving in digital spaces.

This study was motivated by the rapid increase in political polarization observed in Finland over a relatively short period. We examine five topics that have become significantly more polarized online in Finland between 2019 and 2023, a period marked by major global events including a pandemic and a war in Europe. By analyzing polarization trends across two snapshots of the Finnish Twittersphere—centered around the parliamentary elections of 2019 and 2023—we reveal distinct polarization dynamics across different networks. Our findings demonstrate that polarized groups, and the elites and masses within them, shape polarization trends in markedly different ways. This analysis not only provides insight into Finland's evolving political landscape but also offers a method for understanding polarization dynamics in other contexts.

The paper is structured as follows: Section \ref{polarization-measures} briefly reviews current polarization measures. Section \ref{groups-and-hierarchies-in-polarized-networks} outlines our method for inferring hierarchical structures in polarized networks, and decomposes polarization using these hierarchies. Section \ref{hier-pol-res} presents our main findings, analyzing hierarchical polarization trends on Twitter during the 2019 and 2023 Finnish parliamentary elections. We conclude with a discussion in Section \ref{conclusion}.

\section{Polarization Measures}\label{polarization-measures}

Quantifying political polarization in society is a complex task. Traditional approaches have relied on survey-based data to measure opinion bimodalities and issue alignment, estimating ideological shifts and correlations between distinct issue positions (e.g., \cite{kozlowski2021issue}). However, the research landscape has significantly evolved with the emergence of social media platforms, which now serve as critical arenas for daily political discourse and debate. This digital transformation has prompted researchers to increasingly focus on these platforms as rich sources of data for understanding contemporary political dynamics \cite{bright2018explaining, conover2011political, cossard2020falling, green2020covid}.

In social networks, opinion distribution bimodality can be estimated by measuring communication limitations or reachability between opposing groups \cite{morales2015measuring, hohmann2023quantifying}. The underlying assumption is that sharper, more distant peaks in opinion distribution correlate with more isolated communities in topic-specific discussion networks. This isolation is often termed \textit{structural} \cite{SALLOUM2022} or \textit{interactional} \cite{falkenberg2023affective} polarization. Most methods designed to capture this phenomenon assess interaction pattern separation between two distinct groups \cite{GARIMELLA2018}, although recent work has expanded to handle multiple groups \cite{nair2022heterophily, martin2023multipolar}. 

Measuring structural polarization in political communication networks typically involves a three-step process: constructing a network from collected data, identifying functional groups using community detection techniques, and evaluating the division strength with structural polarization scores \cite{GARIMELLA2018, SALLOUM2022}. Common metrics for this final step include Random Walk Controversy \cite{GARIMELLA2018} and Adaptive EI-index \cite{chen2021polarization}, which assess the degree of inter-group isolation or intra-group cohesion. The common intuition these scores hold is to capture the extent to which individuals are ``trapped'' within their own ideological communities or avoid engaging with opposing viewpoints.

Many polarization quantification methods require identifying key network figures, often using endorsement counts (e.g., likes or reshares) as a proxy for importance or ``eliteness'' \cite{GARIMELLA2018, SALLOUM2022, morales2015measuring}. While some studies utilize known node labels to mark politicians in a network \cite{falkenberg2022growing,xia2022russian}, this approach may overlook crucial users without clear affiliations, such as nonpolitical opinion leaders, who often outpace political figures in social media popularity \cite{mukerjee2022political}. Although our method wasn't specifically designed to address this limitation, inferring key figures solely from structural properties can be advantageous in certain scenarios, particularly when explicit labeling is unavailable or insufficient.

Structural polarization scores, while insightful, often have limitations: they typically focus on two-group systems and single-issue interaction patterns, overlooking the critical aspect of issue alignment in multi-group environments. Issue alignment—the degree of agreement across multiple topics—is increasingly recognized as a more pernicious form of polarization \cite{mason2015disrespectfully, chen2021polarization, tornberg2022digital}. It occurs when individuals or groups adopt collective stances on a range of issues based on shared values or ideologies, potentially exacerbating societal divisions \cite{mason2015disrespectfully, tornberg2022digital}. Researchers have applied more classic statistical methods, such as correlation \cite{kozlowski2021issue}, and information theory-based scores for quantifying issue alignment in polarized systems \cite{chen2021polarization}.

\section{Groups and Hierarchies in Polarized Networks}\label{groups-and-hierarchies-in-polarized-networks}

Polarization is often assumed to significantly impact social network structures, giving rise to two internally tightly-knit groups with sparse connections between them \cite{GARIMELLA2018, hohmann2023quantifying}. Measuring polarization separately for elites and the masses requires the inference of their group memberships (for-against) and hierarchy (elite-mass). While from now we will focus on
two polarized groups for this study, this method can be readily adapted to scenarios involving more
than two opposing groups or hierarchies.

We begin by presenting the steps for identifying polarized groups in a network. Next, we partition these groups into smaller hierarchical subgroups based on predefined connectivity patterns and conceptualize their possible interactions in a polarized environment. Finally, we connect the opposing groups, their hierarchies, and actions to existing polarization measures.

\subsection{Identifying the Polarized Groups}

The conventional approach for finding the assortative groups in polarized network is using clustering algorithms \cite{GARIMELLA2018}. We employ \textit{the stochastic block model} (SBM), which is a probabilistic model used to represent and analyze community structure in networks by grouping nodes into blocks, where the probability of connections between nodes depends solely on their block memberships. We specifically use the contrained version of this model, also known as \textit{planted-partition model} \cite{peixoto2014graph_tool, zhang2020statistical, peralta2024multidimensional}, which has been previously applied to finding polarized groups \cite{peralta2024multidimensional}.

The generative nature of SBM enables rigorous model selection, as one can apply \textit{the Occam's razor principle} to select the configuration with lowest description length \cite{grunwald2007minimum}. The property is particularly important as an observed community structure that could be explained due to randomness \cite{guimer2004modul}, can be flagged, as the description length for an SBM with no community structure would be lower than that of a model attempting to overfit by dividing the network into two communities \cite{peixoto2014graph_tool, YAN2016, peixoto2019bayesian, zhang2020statistical}. Therefore, when combined with model selection, SBM-based approach would result in zero polarization in entirely random networks, which is in contrast to most commonly used polarization pipelines introduced in the literature \cite{guimer2004modul, SALLOUM2022}. We sweep the number of blocks (groups) between 1 to 2, and select the configuration leading to the lowest description length. 

\subsection{Unraveling the Hierarchical Structure}

Our second assumption is the existence of hierarchies in each polarized group. Social groups tend to naturally self-organize into hierarchies, where members exhibit varying levels of influence, expertise, or dominance \cite{koski2015understanding,urena2023assortative}. The emergence of these hierarchies is considered both inevitable and often advantageous for social cohesion \cite{koski2015understanding, magee20088}. 

We utilize the well-established concept of core-periphery structures (see e.g. \cite{rombach2017core}) in approximating the hierarchical nature of these polarized groups. The core-periphery structure is a common pattern observed in various systems, including social networks \cite{borgatti2000models, kojakumultiple, YANG201891}, economics \cite{csermely2013structure, wang2016core}, and many other fields \cite{csermely2013structure, kojakumultiple}. It consists of a densely interconnected core, typically representing nodes with higher connectivity, influence, or resource access, surrounded by a sparsely connected periphery representing nodes with lesser influence. In our context, the core can be interpreted as representing the elites, with the periphery acting as the masses. This is in align with the elites typically positioned within structurally dominant areas of networks, forming cohesive structures \cite{motamedi2020examining}. On Twitter, elites often include influential figures such as politicians, journalists, activists, and experts who are active on social media. In contrast, we assume the surrounding periphery comprising the mass, consisting primarily of users who endorse or amplify the core's voices \cite{barbera2015critical}. These assumptions are later assessed for our data in Section \ref{hier-pol-res}. 

For identifying the hierarchical structures that are embedded in the already identified polarized groups, we follow the method proposed by \cite{gallagher2021clarified} to infer the core-periphery structure \cite{gallagher2021clarified} within the SBM framework. We treat each group independently, which enables us to compare their observed hierarchies to the null model separately. We use the common Erd\H{o}s--R\'enyi model as the baseline, as done for instance in \cite{kojaku2018core},  because it assumes no inherent hierarchy, even one that might be explained by the network's degree sequence. This approach has its pros and cons that we address separately in the Appendix \ref{appendix-stat-evidence}. Note that alternative methods for identifying core-periphery structures exist, such as the extended version of the original method \cite{borgatti2000models} for finding multiple core-periphery pairs \cite{kojakumultiple} or the method based on detecting core-periphery structures by surprise \cite{de2019detecting}. 

\subsection{Interactions in Polarized Environments}\label{interactions-in-pol-env}

\begin{figure}[t]
  \centering
  \includegraphics[width=1\linewidth]{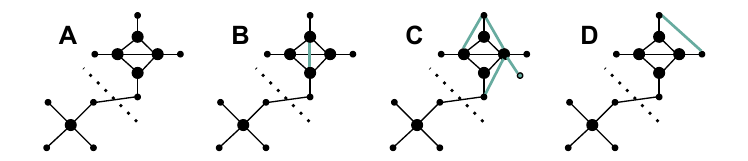}
  \caption{Our conceptualization enables categorizing the actions that lead to an increase in the structural polarization score. \textbf{(A)} is an example of a polarized network that has been partitioned into two groups representing communities with distinct stances on a specific political topic. Large nodes correspond to the elite members, while smaller ones indicate the mass. In \textbf{(B)}, a new connection is formed between two elite members, making the elite group more cohesive. Another polarization increasing action is when a new connection is formed towards the elite group either from an existing node or a new node. This is depicted in \textbf{(C)} and the overall degree of this type of actions is called mass amplification. Lastly, a connection between two members belonging to the mass, is shown in \textbf{(D)}, illustrating the interpretation of mass cohesion. Lower mass cohesion can be seen corresponding to higher centralized opinion leadership among the elites, as most connections are directed towards them.}
  \label{fig:polarization_interactions_plot}
\end{figure}

A common approach is to use a single score to measure the degree of division between these polarized groups (Fig. \ref{fig:polarization_interactions_plot}A). Now, instead of treating all individuals uniformly, we label the strongly internally connected nodes as elites (larger circles), and nodes that have few or weak connections among themselves but well-connected to the central of their group are labeled as the masses (smaller circles). This conceptualization enables us categorize the following interactions:

\textbf{Elite-Elite}: Each new connection between two elites within own group can be deemed as turning the core more connected, and therefore, increasing the \textit{elite cohesion} of that group (Fig. \ref{fig:polarization_interactions_plot}B). 

\textbf{Mass-Elite}: Each elite member can receive an endorsement or validation from the surrounding mass, increasing the total \textit{mass amplification} of the corresponding group in the system (Fig. \ref{fig:polarization_interactions_plot}C).

\textbf{Mass-Mass}: A polarized group may have a very centralized opinion leadership, where mostly elites voices are amplified or more heterogeneous amplification patterns depending on the group's \textit{mass cohesion} (Fig. \ref{fig:polarization_interactions_plot}D), which increases as a connection is formed between two mass members.

\subsection{Defining Structural Polarization and Issue Alignment for Groups and Hierarchies}

Prior work has shown that the most common quantitative methods for measuring structural polarization in social networks perform similarly after they are denoised \cite{SALLOUM2022}. According to the same work, the best performing method in detecting polarized networks was \textit{Adaptive EI-index} (AEI), which is based on the more popular \textit{EI-index} (e.g. \cite{esteve2022political, YAN2016, del2018echo, borah2023impact}) developed by Krackhardt and Stern \cite{krackhardt1988informal}. This method produces a score based on the distribution of connections among nodes within and between groups, where a greater density of links within groups relative to those between groups indicates a more polarized system. 

Let \( G = (V, M) \) represent a network, where \( V \) is the set of nodes and \( M \) is the set of links. We define a partition of \( G \) into two groups, denoted as \( A \) and \( B \), such that: \( V = V_A \cup V_B \) and \( V_A \cap V_B = \emptyset \), where \( V_A \) and \( V_B \) represent the sets of nodes belonging to groups \( A \) and \( B \) respectively. The AEI is defined for such a system as

\begin{equation}
P_{AEI} = \frac{i_{A} + i_{B} - 2 \times e_{AB}}{i_{A} + i_{B}+ 2\times e_{AB}}\,,
\end{equation}
where $i_{X}$ denotes the \textit{internal links density} (observed links divided by the possible number of links) within group $X$ and $e_{AB}$ denotes the \textit{external links density} between groups $A$ and $B$. 

In AEI, all links are treated homogeneously within each group, which prevents it from distinguishing the activity driven by the elites versus the masses. However, one should not view these links equally as the elites are often seen as highly influential individuals on social platforms due to their large following, prominent status in society and access to resources that can amplify their messages and shape public opinion. An elite individual can also connect to other elite members enabling them to reach even a wider audience and have a greater impact on conversations and trends within their communities, making them more cohesive. The amplification and social validation come mainly from the followers, i.e., the surrounding mass, who both take cues from the elites and can be mobilized by them to take specific actions. Therefore, we want to distinguish the link densities according to the hierarchies in the network. We do this by decomposing the elements in $P_{AEI}$ further, such that
\begin{equation}
i_{X} = \frac{I_{c_X} + I_{cp_X} + I_{p_X}}{\frac{1}{2} n_{X} (n_{X} - 1)}\,,
\end{equation}
where $I_{c_X}$ represents the interaction between group $X$'s core nodes, while $I_{cp_X}$ denotes the interaction between its core and periphery, $I_{p_X}$ denotes the interaction among its periphery nodes, and $n_{X}$ denotes the number of nodes in group $X$. We can now substitute the corresponding terms of Eq. (2) into Eq. (1), which gives us

\begin{align}
P_{AEI} &= \overbrace{( \underbrace{\widehat{i}_{c_{A}}}_{\text{elite cohesion}} + \underbrace{\widehat{i}_{cp_{A}}}_{\text{mass amplification}} + \underbrace{\widehat{i}_{p_{A}}}_{\text{mass cohesion}} )}^{\text{Group A's contribution}} \nonumber \\
&\quad + \underbrace{( \, \widehat{i}_{c_{B}} + \widehat{i}_{cp_{B}} + \widehat{i}_{p_{B}} )}_{\text{Group B's contribution}} - \underbrace{2 \, \widehat{e}_{AB}}_{\text{bridge}},
\end{align}

\noindent
where we use the hat symbol to denote the normalized form, where each component is divided by the denominator in Eq. 1.

We refer to Eq. (3) as \textit{polarization decomposition} as it allows us to further study the direct effects of different groups and hierarchies on the polarization score. By splitting the observed polarization score into components, we can detect the differences between groups and their hierarchies on the polarization measure and determine whether their contributions have changed over time. When we say `contribution' or `impact' of a certain group on observed polarization, we specifically refer to these components. This is valuable information in the context of polarization, where we are often interested in monitoring the evolution of the opposing groups and their power dynamics. 

For separating the total contribution of groups from the expected contribution of an individual member of the group, we also measure the \textit{marginal polarization} for both core and periphery nodes. We define marginal polarization as the change in structural polarization when a node of a specific type is added to the network, which enables us to estimate how introducing an average elite user versus an average non-elite user affects the network's overall polarization.

A core node in group $A$ has $k_{c_A}$ links to other core nodes in the same group, it receives $k_{cp_{A}}$ links from the periphery and forms $k_{out}$ external links to group $B$. For a periphery node belonging to group $A$, it can have $k_{p_A}$ links to other periphery nodes. It amplifies $k_{cp_{A}}$ core nodes and connects to $k_{out}$ links outside its own group. By substituting the average of observed values into these decomposed degrees, we can approximate the marginal polarization for a mean core node as

\begin{equation}
\Delta_{c_A} P_{AEI} \approx \frac{2}{\alpha}  \left(\frac{\langle k_{c_A} \rangle + \langle k_{cp_{A}}\rangle }{n_{A}^2} - \frac{\langle k_{out_{B}}\rangle }{n_{A} n_{B}}\right)\,,
\end{equation}
where $\alpha$ denotes the denominator of Eq. 1. The formula is equivalent for the situation where an average periphery node is added to the polarized system $(\Delta_{p_A} P_{AEI}$), but you need to substitute the $k$ values with those reflecting the link distribution of peripheral nodes. Interestingly, the formula suggests that having an equal number of links in and out is not enough to have zero net impact on polarization when group $A$ is smaller than group $B$. See Appendix~\ref{appendix-derivation} for full derivation of this marginal polarization formula.

In addition to structural polarization, we aim to quantify the degree of issue alignment separately for the elites and masses. Issue alignment contributes to political polarization by creating a situation where individuals or groups align themselves strictly along specific issues. When people strongly identify with a particular issue or group and use it as a defining factor for their political stance towards other topics, it can lead to several detrimental consequences \cite{chen2021polarization}. We measure alignment using \textit{the normalized mutual information} (NMI) method  \cite{chen2021polarization} for cores and peripheries separately. In essence, it quantifies the extent to which knowing the group of an user in one topic informs us about the users’s group in another topic.

For estimating each user's stance on a specfic issue, we use their group membership, $A_1$ or $B_1$ for the first network and $A_2$ or $B_2$ for the second network. Since this assessment requires user's presence in both networks, we add the stances for each node in \( V_1 \cap V_2 \) in stance vectors, \( s_1\) and \( s_2 \), corresponding to the respective issues, i.e. these stance vectors encode each user's group membership in the networks. Finally, we compute the issue alignment separately for the distinct hierarchies; NMI(\( s_{1}^{c}\), \( s_{2}^{c}\)) for the cores (elite alignment) and NMI(\( s_{1}^{p}\), \( s_{2}^{p}\)) for the peripheries (mass alignment). This measure is constrained between 0 and 1, where 0 indicates no alignment, and 1 indicates complete alignment between the issues in question.

\section{Hierarchical Polarization in Finnish Parliamentary Elections 2019 and 2023}\label{hier-pol-res}

\begin{figure*}[h]
    \centering
    \includegraphics[width=\linewidth]{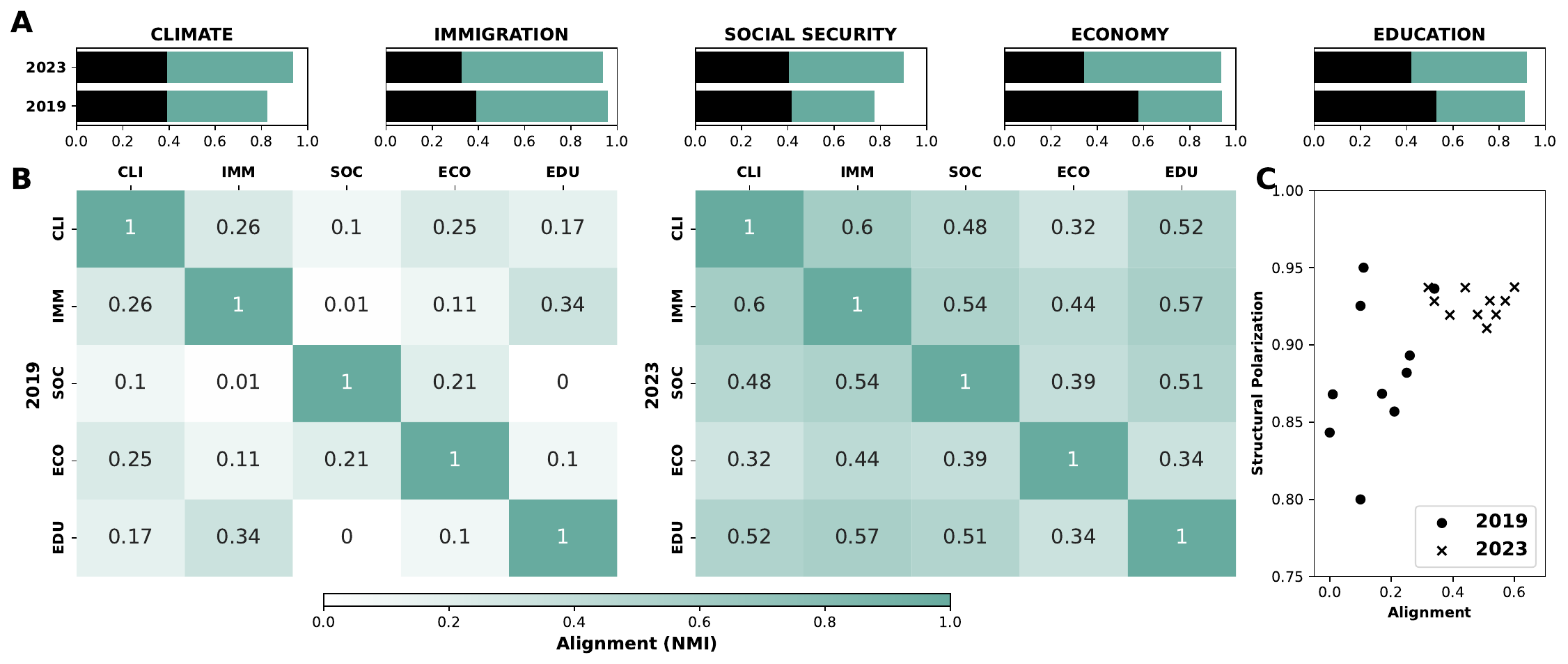}
    \caption{\textbf{(A)} demonstrates the increase in overall structural polarization across all networks, as measured by the AEI. Black bar corresponds to the proportion explained by the null model. The largest increase in the portion not explained by the null model was observed in the network representing economy-related discussions online. \textbf{(B)} The heatmaps depict the evolution of issue alignment over the four years, with 2019 on the left and 2023 on the right. Every pair of topics has experienced a substantial increase in the degree of alignment, as measured by the adjusted NMI. Climate and immigration were already reasonably aligned in 2019, however, the alignment doubled after four years. \textbf{(C)} illustrates the relationship between observed alignment and the average structural polarization scores for all topic pairs in both years—a slight linear relationship appears between the two quantities, possibly indicating that these phenomena are strenghtening each other. Note that in 2019, although some networks had high structural polarization scores, issue alignment remained relatively low. In contrast, by 2023, networks showed both high structural polarization and high issue alignment.}
    \label{fig:overallpol}
\end{figure*}

In this section, we first present the data for our case study, which investigates polarization trends during the Finnish Parliamentary Elections of 2019 and 2023. We demonstrate how the overall structural polarization and issue alignment have increased over four years, and then decompose the polarization scores separately for the elites and the masses.

\subsection{Data}

We use data collected from Twitter (Public API v1 and v2) during the Finnish parliamentary elections in 2019 and 2023. To ensure consistency when comparing these two snapshots, we consider a 12-week period leading up to the election day, resulting in the following periods: from January 21, 2019, to April 14, 2019 for the first elections, and from January 9, 2023, to April 2, 2023 for the second elections. For each year, five networks are constructed from sets of keywords related to larger topics, such as immigration and climate change (see Appendix in \cite{chen2021polarization} for more details). Before constructing the networks, retweets that contain keywords belonging to two or more topics are removed from the dataset, as they would produce a systematic bias, i.e. inflate the extent of alignment between networks. This provides, in a sense, a lower bound for the actual alignment between the networks.

Each network can be viewed as a graph where each user contributing to the topic is assigned to one node. In this graph, an edge between two nodes represents a retweet, which can be deemed as an agreement or a shared point of view on a selected issue between the corresponding users \cite{GARIMELLA2018, chen2021polarization, SALLOUM2022}. The largest connected component is extracted for each network \cite{GARIMELLA2018, chen2021polarization, SALLOUM2022}, with the subsequent removal of self-loops and parallel edges \cite{SALLOUM2022, chen2021polarization}. An average graph consists of approximately 10,000 nodes and 30,000 edges (see Table in Appendix \ref{appendix-networks-summary} for more details).

We tested the assumptions stated in the previous section independently: To verify the presence of political figures within cores, we assessed the probability of a politician being found in the core compared to the periphery. We found that politicians who ran for elections are at least 3-4 times more likely to be in the core than in the periphery (see Appendix \ref{appendix-actors-cores} for more details). To view the connections between the outer periphery and the core nodes as mass amplification, we confirmed this assumption by checking that the majority of retweets flow towards the core nodes (see Appendix \ref{appendix-public-amplification} for more details). Thus, we find our assumptions reasonable.

\subsection{Drastic Increase in Polarization}

\begin{figure*}[t]
    \centering
    \includegraphics[width=\textwidth]{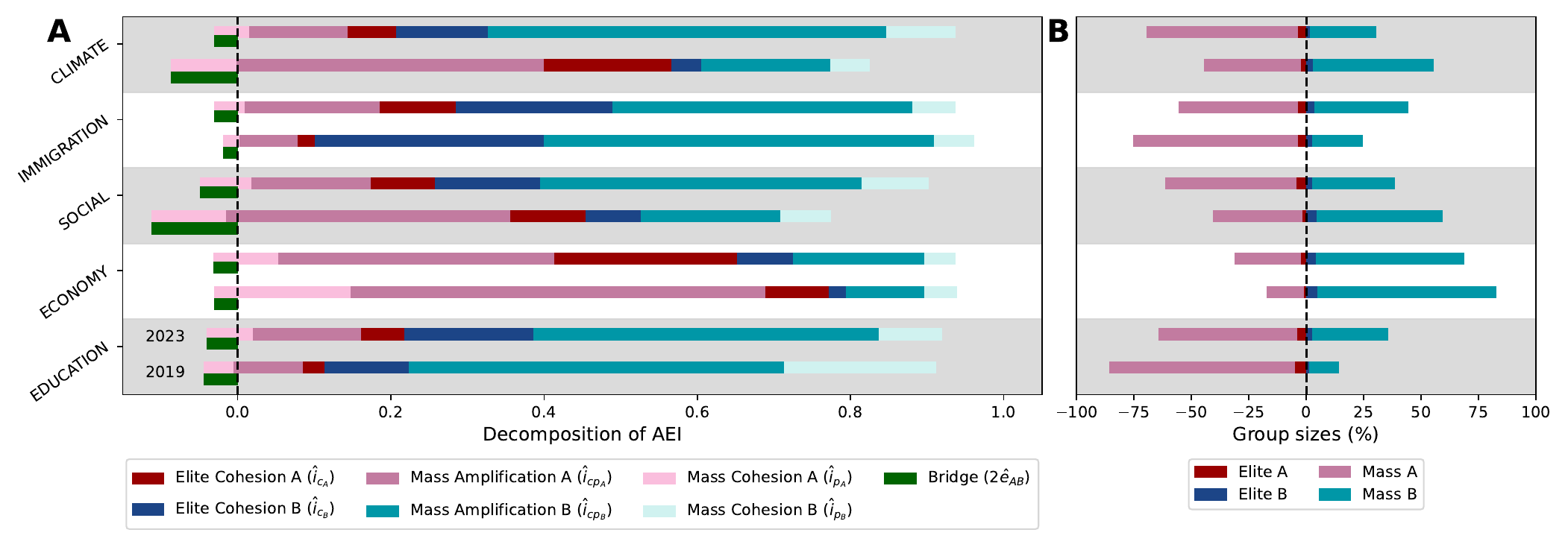}
    \caption{(A) Polarization decomposition for structural contributions of different groups and their hierarchies to AEI-score. The figure illustrates the predominant influence of elite cohesion ($\widehat{i}_{{c}_{A}}$ \& $\widehat{i}_{{c}_{B}}$), mass amplification ($\widehat{i}_{{cp}_{A}}$ \& $\widehat{i}_{{cp}_{B}}$), and mass cohesion ($\widehat{i}_{{p}_{A}}$ \& $\widehat{i}_{{p}_{B}}$) on the overall score. The green part of spectrum represents the impact of the bridge between the opposing entities ($2\times \widehat{e}_{AB}$). The contributions of different hierarchical members not only vary within individual networks but also across the distinct networks. The part of the spectrum that corresponds to the internal structures is shifted to the left by an amount equal to the cross-interactions. This enables us to read the unadjusted AEI score for each network directly from the figure. (B) Groups vary in their sizes, and mostly consists of the masses. Smaller groups can have a great impact on the observed polarization.}
    \label{fig:pol-decomp}
\end{figure*}

Before breaking down polarization by groups and hierarchies, we first report the overall structural polarization and issue alignment across the networks in each snapshot (see Fig. \ref{fig:overallpol}). We confirm that every network passes the assortativity test described in Section \ref{groups-and-hierarchies-in-polarized-networks}.  

Structural polarization has increased in all the networks studied here. Over these four years, climate and economy experienced the highest increase, when taking the null model \cite{SALLOUM2022} into account.
Not only have the opposing groups moved further apart in some individual topics, but they have also become more aligned as the observed alignment values experienced an upward swing in each pair of topics studied here. In other words, the tendency of the same individuals to be divided on other topics as well has become more pronounced. In 2023, immigration aligned the most with the remaining topics, scoring the highest with economy (NMI = 0.6). The least (but substantially) aligned topics appear to be climate and economy, whose issue alignment was already at relatively high level in 2019.

\subsection{Groups have Unequal Impact on the Overall Structural Polarization}

\begin{figure*}[t]
    \centering
    \includegraphics[width=1\textwidth,keepaspectratio]{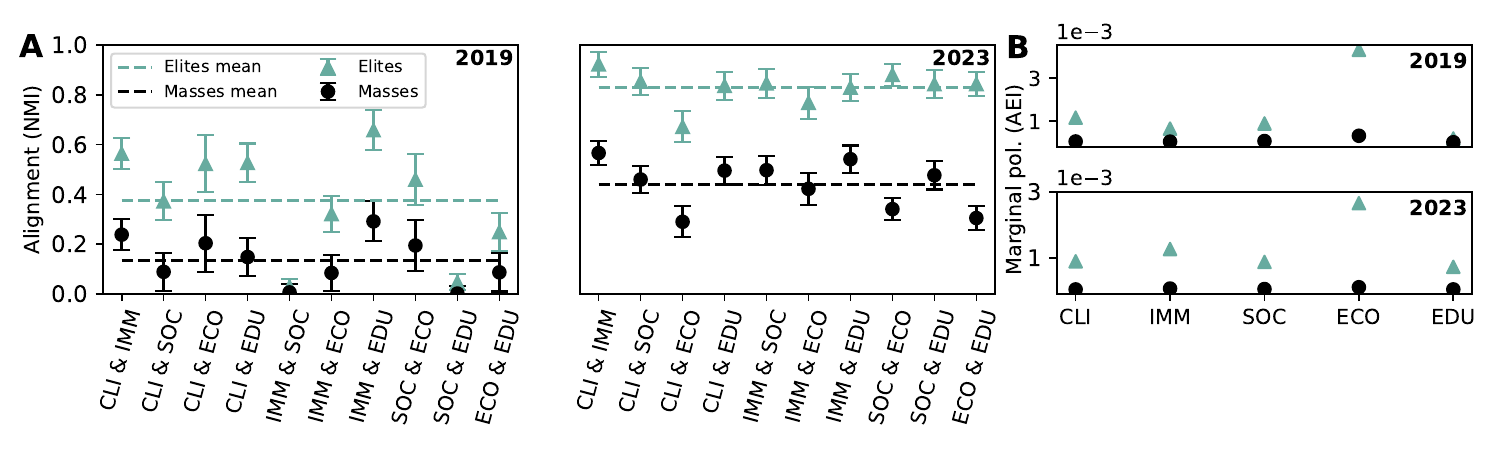}
    \caption{(A) Elites are consistenly more aligned than masses across all topic pairs. Elites became more aligned in 2023, together with a smaller increase in the alignment of mass opinion on various issues. To capture the uncertainty around the observed values, we bootstrapped 500 pairs of networks for each topic pair. Each bootstrap sample represents a subgraph of the original network, where the sizes of cores and peripheries are subject to random fluctuations. We do this by sampling the nodes to groups according to their original group probabilities. (B) Elites tend to have higher marginal polarization as well compared to the mass. In both years, adding a new elite member to the economy network had the greatest impact. A weighted average of both groups' marginal values is applied to obtain a single value representing the hierarchy's mean effect on AEI.}
    \label{fig:summary}
\end{figure*}

Beyond assortativity, we find that each polarized group contains a hierarchy that cannot be explained by the Erd\H{o}s--R\'enyi model. We confirmed assortativity if the two-group model yielded a lower description length than the single-group model. Similarly, hierarchy within each group was confirmed if the core-periphery-model resulted in a lower description length compared to the null model (see Section \ref{groups-and-hierarchies-in-polarized-networks}).

We show the contributions and sizes of different components in Fig. \ref{fig:pol-decomp}, where separate decompositions are created for each year separately for all topics. Here, groups labeled A (in red) refer to left-leaning groups, while groups labeled B (in blue) refer to right-leaning groups, respectively (see Appendix \ref{appendix-bubble-leaning} for labeling details). This allows us to track the changes in the dynamics of left-right groups and their embedded hierarchies in the polarized milieu online. 

Results show many interesting findings. First, the influence of polarized groups on the observed structural polarization varies widely. In fact, in no network was the influence of both groups close to balanced, even when the groups were of similar sizes. In 2019, left-leaning groups contributed the most to the structural polarization in the climate, social security, and economic policy networks, whereas right-leaning groups were more dominant in the immigration and education networks. However, in 2023, right-leaning groups accounted for the majority of the observed polarization in 4 out of 5 networks. The left-leaning group remained the more contributing group in the economic policy network, although its share became slightly smaller. 

The polarization decompositions also reveal the influence and evolution of distinct hierarchical groups. For instance, in the immigration network, the most polarized topic in both years, the elite cohesion of the right-leaning group had a relatively large impact on polarization. The same elite were also significantly more organized than those in the left-leaning group, although the elite cohesion of the latter increased in 2023. The opposite phenomenon was observed for economic policy network, where the left-leaning group's elite cohesion was higher, while right-leaning group elite evolved into more organized structure in 2023. Note how, in every network's observed polarization, the role of at least one elite group's cohesion had increased during the four years; Right-leaning elite became more cohesive in four networks and left-leaning elite in three networks, indicating a stronger role of elite in certain topics. The most drastic swap in terms of contributions of a hierarchical groups was witnessed in the mass amplification patterns in climate network, where right-leaning public outrun the previously dominating left-leaning public. Although in the absolute terms, the publics' interaction patterns tend to explain most of the observed polarization in all networks, the relative contribution (when compared to the sizes of these hierarchical groups) of the elite members is, on average, greater, which we see in Fig. \ref{fig:pol-decomp}B and later in Fig. \ref{fig:summary}B.

Communication between polarized groups has mostly dropped or remained at the same level across all topics. This type of communication, labeled as bridge, is the only interaction that could structurally decrease polarization. Its share of the decomposition has shrunk in all networks except the immigration network.

Three main findings from the Fig. \ref{fig:pol-decomp} stand out: first, the increase in structural polarization does not follow uniform patterns, second, a larger group size does not always lead to a higher contribution to the observed polarization, and third, the elites play relatively higher role than the masses. We see, for instance, how right-leaning group compensated its smaller size in climate and education with more organized core and greater mass amplification in 2023, leading to substantially higher impact on polarization when compared to the impact of left-leaning group. The greater relative contribution of the elite members is generally evident across all networks when reflecting it to the size of their groups shown in Fig. \ref{fig:pol-decomp}B. We also confirm this by computing the marginal polarization separately for the elite and mass individuals in Fig. \ref{fig:summary}B.

\subsection{Elites are Significantly More Aligned than the Masses}

Elite members are consistently more aligned than the public in all topics studied here, as shown in Fig. \ref{fig:summary}A. Higher alignment suggests that the elite regularly maintain more predictable stances on various political themes. In 2019, among both elites and masses, topic pairs immigration-education and immigration-climate were the most aligned. Conversely, the lowest alignment was found between the topics immigration-social security and education-social security. 

After four years the gap between elite and mass alignment widened—not only did the elite show a much higher level of issue alignment (the mean level doubled), but the disparity between the masses and elites also increased in 2023. The overall level of mass alignment did not stay at the same level as in 2019 either. Instead, the mass also experienced a notable increase in alignment across all topics, as the mean line, corresponding to the average of alignment over all the topics, rose from 0.13 to 0.44. For the elite groups, eight out of ten pairs of topics had an NMI value over 0.8, which can be considered very high. Results show that knowing elite member's stance on climate politics would reveal the most information on their views about immigration policies compared to the other topics included in this study. Interestingly, climate-economy experienced the smallest increase in alignment.

\subsection{Elites and Masses of Opposing Groups Show Distinct Activity Patterns}

\begin{figure*}
    \centering
    \includegraphics[width=1\textwidth,keepaspectratio]{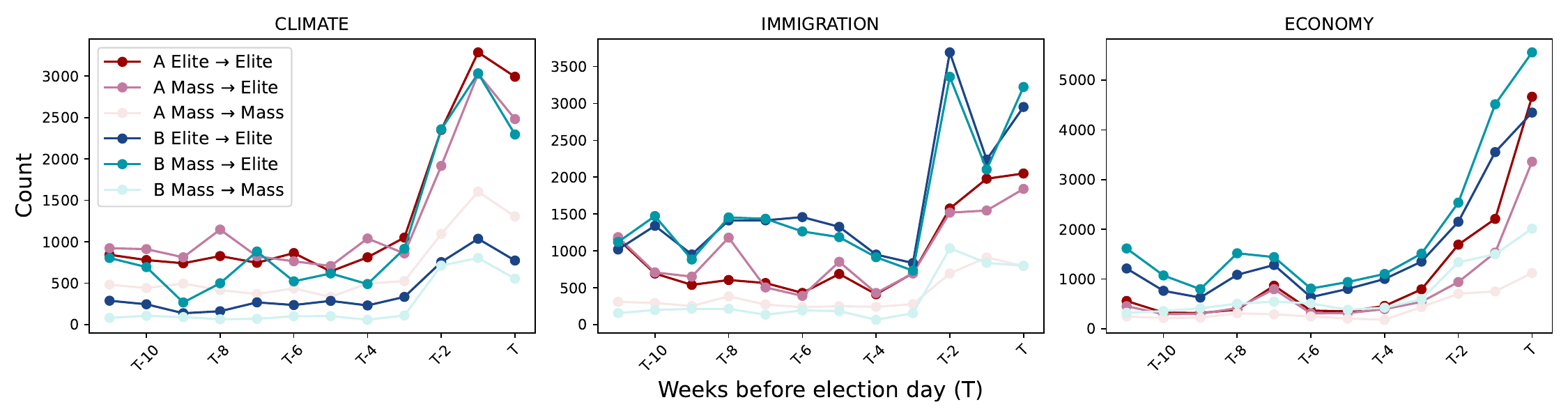}
    \caption{Activity patterns in the most polarized networks in 2023 separated at group and hierarchy level. In all networks, largest jump in activity takes place approximately three weeks before the election day, particularly within right-leaning group in immigration network. Which opposing group is more active depends on the issue. For instance, activity within the right-leaning elites is substantially higher in immigration and economy, whereas in climate, social security and education, left-leaning elites appear to be more active. The extent of activity of a specific group does not translate into the observed polarization. Figures for remaining topics and for 2019 can be found in Appendix \ref{appendix-activity-patterns}.}
    \label{fig:trends}
\end{figure*}

So far, we have focused on aggregate networks, meaning we constructed networks based on all interactions that occurred during the specified time period. These networks were unweighted, as there are no established guidelines for computing polarization in weighted graphs (or whether it should be done at all).

A polarized group might exhibit high cohesion among elites or the masses, yet show non-uniform endorsement patterns, suggesting that once-formed ties are not consistently maintained. Similarly, peaks in mass amplification may occur as bursts, while otherwise following stable patterns. To assess the temporal dynamics of interactions between elites and masses, we labeled each interaction according to the identified groups and their hierarchies (see Section \ref{interactions-in-pol-env}), and plotted them leading up to the election day (see Fig. \ref{fig:trends}).

The most crucial finding is that a higher level of activity does not dictate group's structural contribution to the observed polarization. For instance, in the climate network, the elite cohesion of the right-leaning group was approximately twice as large as that of the left-leaning elite in 2023. Yet, left-leaning elites consistently endorsed each other more than right-leaning elites throughout the entire time period. However, in immigration network, where right-leaning group also dominated in terms of elite cohesion and mass amplification in 2023, also showed more active elites and masses. 

For the most polarized networks in 2023 (climate, immigration, and economy), it is apparent that the activity of elite groups varies depending on the issue. In some topics, such as immigration and economy, mass-elite and elite-elite activity are closely matched for both groups. In others, like climate, mass-elite interactions can substantially exceed elite-elite interactions, particularly within the right-leaning group.

\section{Discussions and Conclusions}\label{conclusion}

This study examined the differential influence of polarized groups and their hierarchies on online polarization. We developed a method to identify these structures within a single social system and analyzed their impacts on structural polarization and issue alignment-based polarization. Applying this method to Twitter networks during Finnish parliamentary elections yielded revealing insights. 

Our findings demonstrated an unequal impact of polarized groups on overall structural polarization across all studied networks, challenging the notion of balanced systems where opposing groups contribute equally to polarization scores. In 2019, left-leaning groups dominated in three out of five topics (climate, social security, and economy). However, by 2023, right-leaning groups became the dominant force in climate and social security networks, primarily due to a more organized elite and greater mass amplification. These shifts align with Finland's political transitions: a left-leaning government in 2019 and a right-leaning one in 2023 \cite{samforsk2023}.

User hierarchy in polarized networks is crucial. Our findings reveal that elites consistently demonstrate higher alignment on distinct political topics and exert relatively greater influence on structural polarization compared to the masses. Networks experiencing increased overall polarization invariably showed at least one elite group with enhanced cohesion, underscoring that heightened polarization doesn't necessarily require greater mass amplification. The right-leaning group in the education network exemplifies how elite organization alone can increase polarization. We also observed how elite and mass alignment more than doubled over the four years. The relative increase in issue alignment was more pronounced for the masses (+228\%) than for the elites (+121\%), suggesting that the public has not remained insulated from the trend towards increased alignment.

We also examined activity patterns among elites and masses, as well as between opposing groups. Our findings revealed that high activity levels in certain subgroups do not necessarily correlate with increased observed polarization. While extending current polarization scores to account for user activity is possible, the optimal method for doing so remains unclear in polarization literature. Consequently, we analyzed group and hierarchical activities separately, proving the complexities we have in quantifying political polarization.

Our work helps make sense of today's polarized environments. The overrepresentation of elite voices in traditional and social media may skew public perception of polarization, potentially exacerbating feelings of division. We provide approaches to precisely quantify the degree of this phenomenon by comparing elite polarization to that of the surrounding public in online spaces. Furthermore, by treating elite and mass polarization as distinct processes, our method enables the detection of anomalous patterns, such as unexpected surges in mass amplification. 

The significance of hierarchies and their impacts on online systems should not be underestimated. Our findings enable the deeper exploration into these matters, enriching the comprehension of mechanisms that drive polarization.

\bibliographystyle{acm}
\bibliography{main}

\appendix

\section{Marginal Polarization}\label{appendix-derivation}

How much does the overall structural polarization increase if you added an ``average'' core (or periphery) node to group A? Assume that  an average core node has $k_{c_A}$ links to other core nodes in the same group, it receives $k_{cp_{A}}$ links from the periphery and forms $k_{out}$ links to the other group B. The marginal polarization could be written then as

\begin{align}
\Delta_{c_{A}} P_{AEI}  &= 
\frac{
\frac{(I_{c_{A}} + k_{c_A} + I_{cp_{A}} + k_{cp_{A}} + I_{p_{A}})}{\binom{n_{A} + 1}{2}} 
+ 
\frac{(I_{c_{B}} + I_{cp_{B}} + I_{p_{B}})}{\binom{n_{B}}{2}}
}{\alpha'} \nonumber \\
&\quad - 2 \times \frac{E + k_{out_{B}}}{(n_{A} + 1) n_{B}} \nonumber \\
&\quad - 
\frac{
\frac{(I_{c_{A}} + I_{cp_{A}} + I_{p_{A}})}{\binom{n_{A}}{2}} 
+ 
\frac{(I_{c_{B}} + I_{cp_{B}} + I_{p_{B}})}{\binom{n_{B}}{2}} 
}{\alpha} \nonumber \\
&\quad - 2 \times \frac{E}{n_{A} n_{B}}\,, 
\end{align}

where  $I_{\bullet}$ is the observed number of links within the corresponding structure, and $E$ denotes the observed number of links between groups $A$ and $B$. Each term is adjusted for density, as described in the main text. We approximate $\alpha' = \alpha$. Hence, we get for average core node the following marginal:

\begin{align}
\Delta_{c_{A}} P_{AEI}  &= 
\frac{
\frac{(I_{c_{A}} + k_{c_A} + I_{cp_{A}} + k_{cp_{A}} + I_{p_{A}})}{\frac{1}{2}n_{A}(n_{A}+1)} 
- 2 \times \frac{E + k_{out_{B}}}{(n_{A} + 1) n_{B}} 
}{\alpha} \nonumber \\
&\quad - 
\frac{
\frac{(I_{c_{A}} + I_{cp_{A}} + I_{p_{A}})}{\frac{1}{2}n_{A}(n_{A}-1)}
+ 2 \times \frac{E}{n_{A} n_{B}}
}{\alpha} \nonumber \\
&\approx
\frac{
\frac{(I_{c_{A}} + k_{c_A} + I_{cp_{A}} + k_{cp_{A}} + I_{p_{A}})}{\frac{1}{2}n_{A}(n_{A})} 
- 2 \times \frac{E + k_{out_{B}}}{n_{A} n_{B}} 
}{\alpha} \nonumber \\
&\quad - 
\frac{
\frac{(I_{c_{A}} + I_{cp_{A}} + I_{p_{A}})}{\frac{1}{2}n_{A}(n_{A})} 
+ 2 \times \frac{E}{n_{A} n_{B}}
}{\alpha} \nonumber \\
&= 
\frac{
\frac{( k_{c_A} + k_{cp_{A}})}{\frac{1}{2}n_{A}(n_{A})} 
- 2 \times \frac{k_{out_{B}}}{n_{A} n_{B}}
}{\alpha} \nonumber \\
&= 
2 \times \frac{
\frac{( k_{c_A} + k_{cp_{A}})}{(n_{A})^2} 
- \frac{k_{out_{B}}}{n_{A} n_{B}}
}{\alpha}
\end{align}

Similarly, for the remaining node types:

\begin{align}
\Delta_{c_{B}} P_{AEI} &\approx \frac{2}{\alpha} \times (\frac{( k_{c_B} + k_{cp_{B}})}{n_{B}^2} - \frac{k_{out_{A}}}{n_{B} n_{A}}) \\
\Delta_{p_{A}} P_{AEI} &\approx \frac{2}{\alpha} \times (\frac{( k_{p_A} + k_{cp_{A}})}{n_{A}^2} - \frac{k_{out_{B}}}{n_{A} n_{B}}) \\
\Delta_{p_{B}} P_{AEI} &\approx \frac{2}{\alpha} \times (\frac{( k_{p_B} + k_{cp_{B}})}{n_{B}^2} - \frac{k_{out_{A}}}{n_{B} n_{A}})
\end{align}

\section{Statistical Testing}\label{appendix-stat-evidence}

Our method requires the network to be partitioned into $k$ polarized groups (assortativity assumption) after which each polarized group is decomposed further into separate hierarchical groups (hierarchy assumption). Therefore, in case of $k=2$, the overall number of different groups would be four. From now, we assume $k=2$, and label the identified polarized groups as $A$ and $B$. For the hierarchical decomposition, we label the cores of $A$ as $c_{A}$ and its periphery as $p_{A}$ (similary for B).

We know that most existing methods for determining the modular structure of networks face issues. They often overlook the statistical evidence supporting the identified patterns, making it difficult to distinguish genuine structures from noise. In our case, we want to be sure about two things: 1) Is there enough evidence that the network can be decomposed into two groups? \& 2) If so, is there enough evidence that these groups can be partitioned further into separate hierarchies? 

Each network undergoes the following partitioning pipeline: 1) Network is partitioned into two blocks with the non-uniform planted partition model. 2) To find the embedded hierarchies, we consider the groups independently, and run the core-periphery algorithm provided in \cite{gallagher2021clarified}. We apply the traditional hub-and-spoke description of core-periphery structure as it has been shown to explain online amplification better than the layered one. We run this two-step decomposition for 100 times and save each obtained partition for the statistical significance testing.

We want to be explicit about how we measure significance and what we do with that information. First, if it turns out that the network has no significant assortative groups, then we conclude that there is no evidence for the network being at least structurally polarized. In our experiment, all networks were better described with two groups in terms of description length. We then shift our attention to the inferred hierarchies within each group. 

First, defining a hierarchy is harder than defining the absence of it. Perhaps the most common baseline model in network science, Erd\H{o}s--R\'enyi model, generates random graphs, where each edge $(i,j)$ is included in the graph with probability $p$ independently from every other edge. Hence, the possible heterogeneity that emerge in such graph is purely due to random fluctuations, making it decent candidate for null model.

A stricter null model would be the configuration model, another common option for testing the statistical evidence of an observed structure. Essentially, it is a mathematical model used to generate random networks with a specified degree sequence. Interestingly, previous research has shown how the degree sequence of a network can fully explain the core-periphery structure in certain situations \cite{kojaku2018core}. We selected ER-model over the configuration model for three reasons:

\begin{enumerate}
    \item We use the core-periphery structure to identify the structurally dominant users, and it is acceptable for this significance to stem from a user's higher degree (as in e.g. \cite{GARIMELLA2018}). 

    \item Some work (e.g. \cite{10.1214/23-SS141}) has discussed the possible shortcomings of using configuration model in detecting core-periphery structures, claiming it likely leading to low power for the statistical test. They illustrate this with an ideal core-periphery structure, for which the same original network is the only one that maintains the fixed degree sequence. Therefore, no rewiring is possible, resulting in a hypothesis test yielding a p-value of 1. \cite{kojaku2018core} claims that an additional block, such as a community or another core-periphery, is required to conclude the significance of the observed core-periphery. However, as far as we know, there isn't a generative model capable of inferring multiple core-periphery structures simultaneously, and such model is typically required to measure the description length.

    \item Erd\H{o}s--R\'enyi model is an intuitive and simple model to understand, and it does not produce self-loops or parallel edges.
    
\end{enumerate}

Ideally, we would apply directly a constrained version of nested stochastic block model \cite{PhysRevX.4.011047} because of the hierarchical nature of our overall decomposition approach. Unfortunately, it is out of scope of this work to develop such method, and hence we employ the method described in the main text for model selection.

\section{Summary of Networks}\label{appendix-networks-summary}

\begin{table}[h]
    \centering
    \caption{Retweet networks were constructed for both election years, covering five distinct networks each. These networks were constructed based on Twitter data obtained over a span of 12 weeks leading up to the respective election day, which were 14.4. for 2019 and 2.4. for 2023. $|N|$ represents the count of unique nodes (users), and $|E|$ denotes the count of unique edges (retweets) in the network after the preprocessing.}
    \label{tab:networks-summary}
    \begin{tabular}{c|cc|cc}
        \toprule
        Topic & $|N|_{19}$ & $|E|_{19}$ & $|N|_{23}$ & $|E|_{23}$ \\
        \midrule
        Climate politics & 16 639 & 50 605 & 11 264 & 34 509 \\
        Immigration politics & 6 739 & 20 923 & 9 519 & 36 889 \\
        Social security & 13 926 & 39 778 & 11 395 & 34 173 \\
        Economic policy & 5 880 & 10 962 & 10 031 & 37 111 \\
        Education policy & 9 843 & 20 314 & 12 814 & 35 619 \\
    \bottomrule
\end{tabular}
\end{table}

\section{Political Actors in Cores versus Peripheries}\label{appendix-actors-cores}

To verify that the presence of political elite is higher in the cores than in peripheries, we computed the following for each network: How much more likely it was for a randomly selected node from the cores to belong to the list of political candidates compared to a node sampled from the peripheries, i.e. 

\[
\gamma = \frac{\text{Probability of core node being a candidate}}{\text{Probability of periphery node being a candidate}}
\]

We report the values of $\gamma$ for each network in Table \ref{tab:topics}.

\begin{table}
    \centering
    \caption{Values of $\gamma$ demonstrate the higher chance of finding a politician from the cores versus peripheries.}
    \begin{tabular}{ll}
        \toprule
        \textbf{Topic} & $\gamma$ \\
        \midrule
        CLIMATE & 3.3\\ 
        IMMIGRATION & 2.8\\
        SOCIAL SECURITY & 3.1 \\
        ECONOMIC POLICY & 4.0 \\
        EDUCATION & 3.6 \\
        \bottomrule
        \\
    \end{tabular}
    \label{tab:topics}
\end{table}

\section{Public Amplification}\label{appendix-public-amplification}

\begin{figure*}[h!]
    \centering 
    \includegraphics[width=0.7\textwidth]{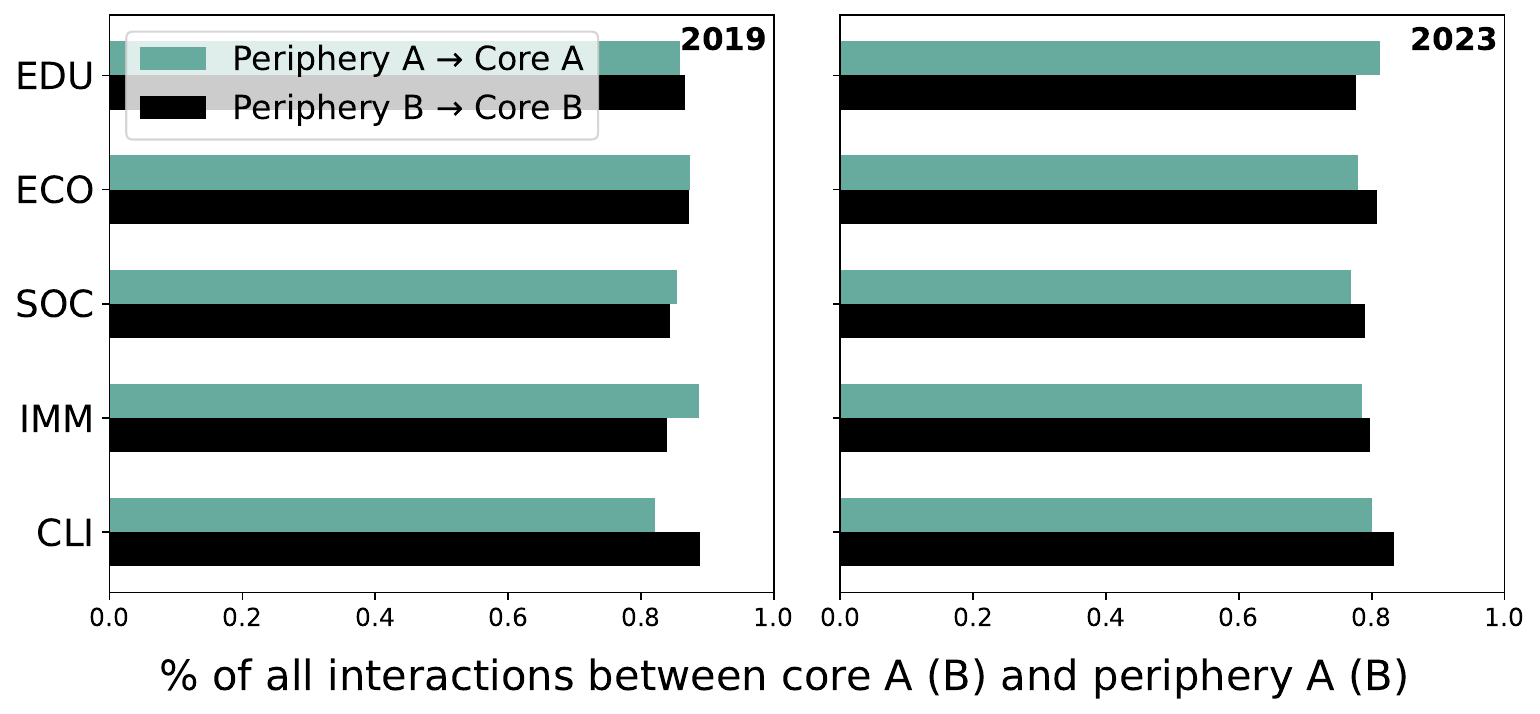} 
    \caption{To view the core-periphery interactions as mass amplification, we confirm that the majority of connections are directed towards the core. In all networks examined, most of the links between the core and periphery consist of retweets originating from the outer periphery. This proportion remains relatively consistent across different polarized groups. Therefore, it's reasonable to characterize this dynamic as mass amplification.}
    \label{fig:figure-public-amplification}
\end{figure*}

\section{Political Parties in Inferred Polarized Groups}\label{appendix-bubble-leaning}

\begin{figure*}[h!]
    \centering 
    \includegraphics[width=\textwidth]{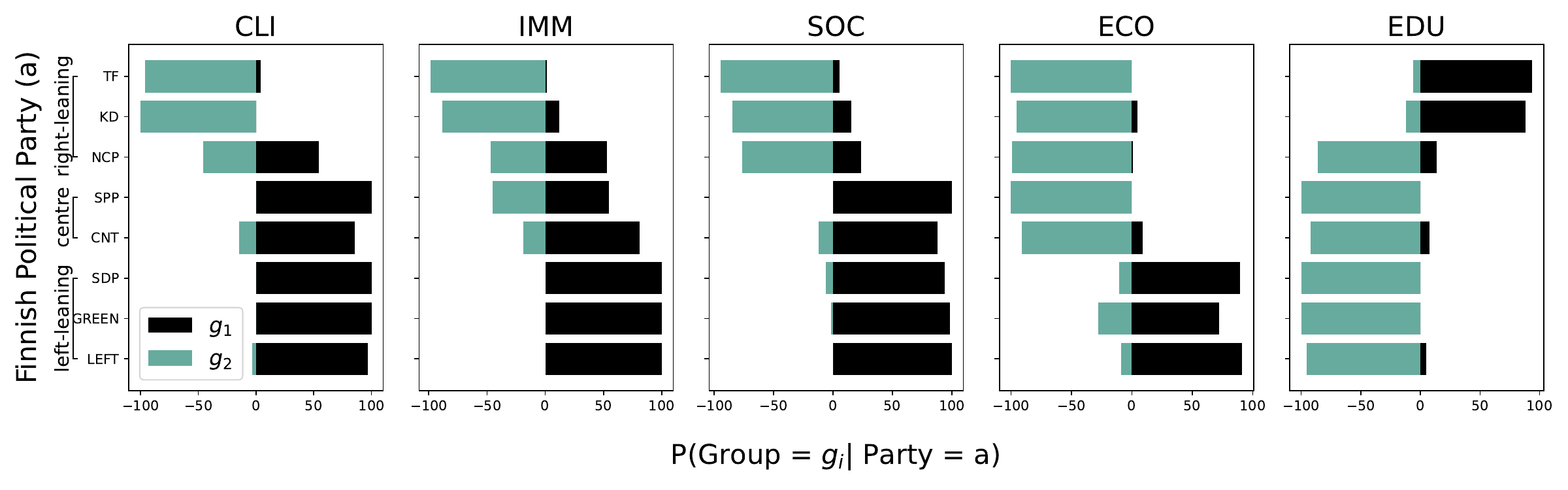} 
    \caption{Majority of the candidates from left-leaning parties are grouped together in the inferred polarized groups, as are those from right-leaning parties.}
    \label{fig:appendix-party-distribution}
\end{figure*}

We verified that the inferred groups can be called either left-leaning or right-leaning based on the real political affiliations of the political actors in the groups in Fig. \ref{fig:appendix-party-distribution}.

\section{Activity Patterns}\label{appendix-activity-patterns}

\begin{figure*}[ht!]
    \centering
    \includegraphics[width=\textwidth]{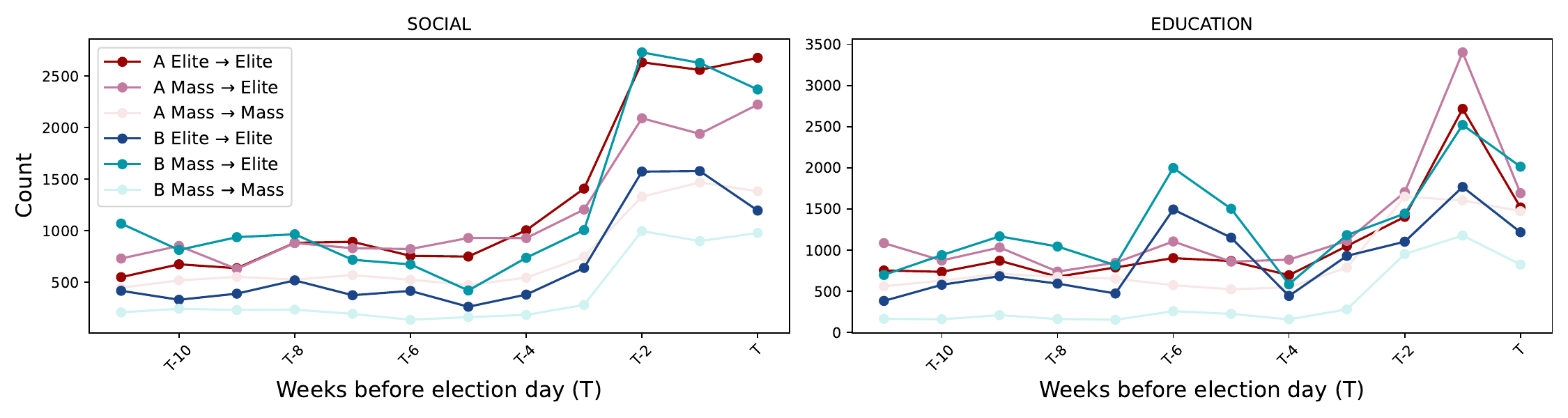}
    \caption{Activity patterns in social security and education networks 2023. See caption of Figure \ref{fig:trends} for more details.}
    \label{fig:trends1}
\end{figure*}

\begin{figure*}[h!]
    \centering
    \includegraphics[width=\textwidth]{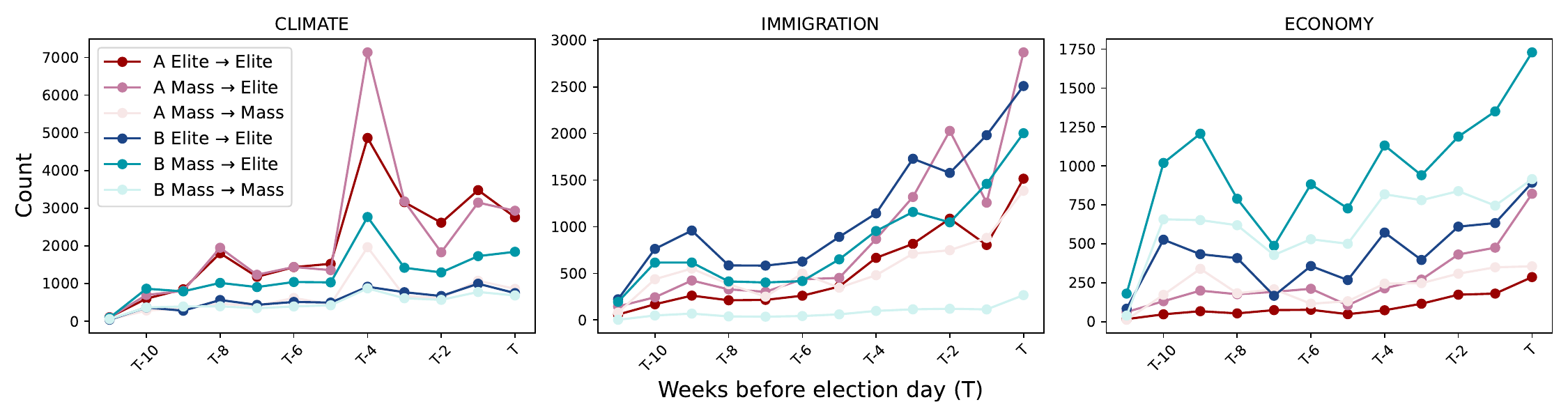}
    \caption{Activity patterns in climate, immigration and economy networks 2019. See caption of Figure \ref{fig:trends} for more details.}
    \label{fig:trends2}
\end{figure*}

\begin{figure*}[h!]
    \centering
    \includegraphics[width=\textwidth]{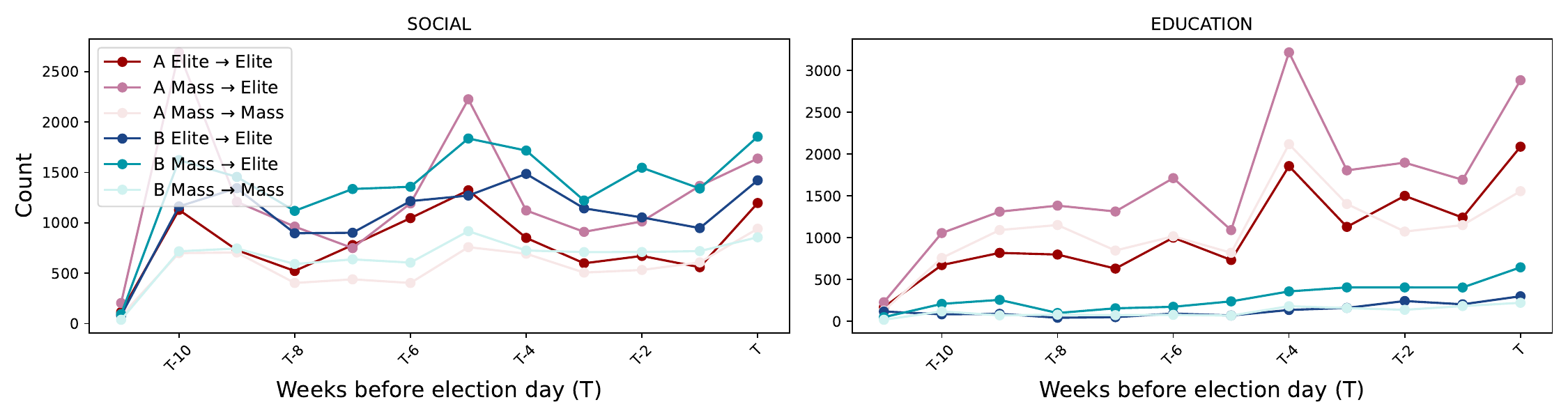}
    \caption{Activity patterns in social security and education networks 2019. See caption of Figure \ref{fig:trends} for more details.}
    \label{fig:trends3}
\end{figure*}

\end{document}